\documentclass[12pt,preprint]{aastex}

\slugcomment{Printed \today}

\shorttitle{Single Disk in SVS 13 Binary System}
\shortauthors{Anglada et al.}

\begin{document}

\title{A Single Circumstellar Disk in the SVS 13 Close Binary System}

 \author{Guillem Anglada\altaffilmark{1}, Luis F. Rodr\'\i
guez\altaffilmark{2}, Mayra Osorio\altaffilmark{1}, Jos\'e M.  
Torrelles\altaffilmark{3}, Robert Estalella\altaffilmark{4,5}, Maria T.
Beltr\'an\altaffilmark{6}, Paul T.  P. Ho\altaffilmark{5,7}}

 \altaffiltext{1}{Instituto de Astrof\'\i sica de Andaluc\'\i a, CSIC,
Camino Bajo de Hu\'etor 24, E-18008 Granada, Spain;
 {guillem@iaa.es, osorio@iaa.es}}

 \altaffiltext{2}{Centro de Radioastronom\'{\i}a y Astrof\'{\i}sica, UNAM,
Apartado Postal 3-72 (Xangari), 58090 Morelia, Michoac\'an, M\'exico;
 {l.rodriguez@astrosmo.unam.mx}}

 \altaffiltext{3}{Instituto de Ciencias del Espacio (CSIC)-IEEC, Gran
Capit\`a, 2, 08034 Barcelona, Spain;
 {torrelles@ieec.fcr.es}}

 \altaffiltext{4}{Departament d'Astronomia i Meteorologia, Universitat de
Barcelona, Av. Diagonal 647, E-08028 Barcelona, Spain;
 {robert@am.ub.es}}

 \altaffiltext{5}{Harvard-Smithsonian Center for Astrophysics, 60 Garden
Street, Cambridge, MA 02138, USA;
 {ho@cfa.harvard.edu}}

 \altaffiltext{6}{Osservatorio Astrofisico di Arcetri, Largo E. Fermi 5,
I-50125 Firenze, Italy;
 {mbeltran@arcetri.astro.it}}

\altaffiltext{7}{Academia Sinica Institute of Astronomy and Astrophysics, 
Taipei, Taiwan}

\begin{abstract}

We present Very Large Array observations at 7 mm of the sources IRAS 2A,
IRAS 2B, MMS2, MMS3 and SVS 13, in the NGC1333 region. SVS 13 is a young
close binary system whose components are separated by 65 AU in projection.
Our high angular resolution observations reveal that only one of the
components of the SVS 13 system (VLA 4B) is associated with detectable
circumstellar dust emission.  This result is in contrast with the well
known case of L1551~IRS5, a binary system of two protostars separated by
45 AU, where each component is associated with a disk of dust. Both in
SVS 13 and in L1551~IRS5 the emission apparently arises from compact
accretion disks, smaller than those observed around single stars, but
still massive enough to form planetary systems like the solar one. These
observational results confirm that the formation of planets can occur in
close binary systems, either in one or in both components of the system,
depending on the specific angular momentum of the infalling material.

\end{abstract}

\keywords{ISM: jets and outflows --- ISM: individual (NGC 1333, SVS 13, HH
7-11) --- radio continuum: ISM --- stars: formation}

\section{INTRODUCTION}

SVS~13, in the NGC1333 region, was discovered as a 2.2 $\mu$m source by
Strom, Vrba, \& Strom (1976), and since the source is roughly aligned with
the chain of Herbig-Haro objects 7-11 (Herbig 1974; Strom, Grasdalen, \&
Strom 1974), it was assumed to be the exciting source of this classical HH
system. Later, Goodrich (1986) detected a faint visible counterpart of
SVS~13. However, the star SVS~13 presents a number of peculiar properties.
The source exhibited a significant increase of its brightness at optical
($\sim$3 mag), and IR ($\sim$1 mag) wavelengths in 1988-1990 (Eisl\"offel
et al. 1991; Liseau, Lorenzetti, \& Molinari 1992; Harvey et al. 1998),
and since then, the flux has remained almost steady (Aspin \& Sandell
1994; Khanzadyan et al. 2003). In addition, despite being optically
visible, indicating that it is a relatively evolved young object, SVS 13
is a strong millimeter source, known as MMS1 (e.g., Grossman et al. 1987;
Looney, Mundy, \& Welch 2000), and presents other characteristics, such as
the presence of an extremely high velocity CO outflow, that suggest it is
in a much earlier evolutionary stage (a Class 0/I object; Bachiller et al.
2000).

Radio continuum emission from SVS 13 was first reported by Snell \& Bally
(1986). Rodr\'{\i}guez, Anglada, \& Curiel (1997, 1999) mapped SVS 13 with
the Very Large Array (VLA) at 3.6 and 6 cm (their source VLA 4).  
Anglada, Rodr\'{\i}guez, \& Torrelles (2000), through VLA observations at
3.6 cm of higher angular resolution and sensitivity, discovered that
SVS~13 is, in fact, a close binary system. The two components of the
binary (VLA 4A and VLA 4B) are separated by $0\rlap.''3$, corresponding to
65 AU in projection (assuming a distance of 220 pc; \v{C}ernis 1990), and
have similar flux densities at 3.6 cm.  The water masers associated with
SVS~13 appear segregated in position and velocity, supporting the binary
hypothesis (Rodr\'{\i}guez et al. 2002).  Anglada et al.  (2000) noted
that the optical position for SVS~13 (as measured by Rodr\'{\i}guez et al.
1997) is closer to VLA~4A, while the millimeter position (Looney et al.
2000) is closer to VLA~4B.  Although the precision of the astrometry
available at that time did not allow an unambiguous association, this
result led these authors to suggest that the strong millimeter emission
reported for SVS~13 could arise from only one of the components of the
binary (VLA~4B, the eastern component), while the optical emission would
come from the other component (likely from VLA~4A, the western component).  
In the interpretation proposed by Anglada et al. (2000), only one of the
stars (VLA 4B) is surrounded by a dusty envelope or disk, while the other
(VLA 4A, the visible star) is not. A similar interpretation has been
proposed recently by Loinard et al. (2002) for IRAS 04368+2557 in L1527,
on the basis of the morphology of the two obscured sources observed at 7
mm.

A confirmation of this interpretation for SVS 13 requires a precise
comparison of the positions of the sources observed at different
wavelengths, in order to identify the individual contribution of each
component of the binary. Since the angular separation between VLA 4A and
VLA 4B is only $0\rlap.''3$, absolute astrometry down to $<0\rlap.''1$ is
required in order to obtain an accurate enough registration of the
positions. Although the accuracy of the absolute astrometry of the optical
observations ($\pm0\rlap.''3$) is difficult to improve due to the lack of
a large enough number of suitable reference stars in the field, it is
possible to improve the accuracy of the registration between the
centimeter and millimeter positions by using the same instrument and
calibration procedures in both wavelength ranges.

In this Letter, we present VLA observations at 7 mm of the region near
SVS 13, carried out in the D and B configurations, using the same phase
calibrator and procedures as in the previous VLA observations at 3.6 cm.  
The B configuration observations at 7 mm provide an angular resolution of
$\sim0\rlap.''2$, similar to that of the A configuration at 3.6 cm, and an
expected accuracy in the registration between both images down to
$<0\rlap.''05$, allowing a precise comparison of the emission at both
wavelengths, necessary to test the single disk hypothesis for the SVS 13
binary. These observations also provide 7 mm data on other sources in
NGC1333:  VLA 2 (MMS3), VLA 7 (IRAS 2A), VLA 10 (IRAS 2B), and VLA 17
(MMS2).

\section{OBSERVATIONS}

The observations were carried out at 7 mm using the VLA of the National
Radio Astronomy Observatory (NRAO)\footnote{NRAO is a facility of the
National Science Foundation operated under cooperative agreement by
Associated Universities, Inc.}\ in the D and B configurations.  The D
configuration observations were carried out during 2000 September 7 using
two phase centers, one near SVS 13 (at the position $\alpha (\rm J2000) =
03^h 29^m 03\rlap.{^s}57$; $\delta(\rm J2000) = +31^\circ 16' 02\farcs8$)
and a second near the source IRAS 2 (at the position $\alpha (\rm J2000) =
03^h 28^m 56\rlap.{^s}14$; $\delta(\rm J2000) = +31^\circ 14' 31\farcs2$).
For all observations an effective bandwidth of 100~MHz with two circular
polarizations was employed.  The absolute amplitude calibrator was
1331+305 (adopted flux density of 1.45 Jy) and the phase calibrator was
0336+323, with a bootstrapped flux density of 2.1$\pm$0.1 Jy.  The data
were edited and calibrated using the software package Astronomical Image
Processing System (AIPS) of NRAO.  Cleaned maps were obtained using the
task IMAGR of AIPS and natural weighting. The resulting synthesized beam
size was $2\rlap.''0\times1\rlap.''6$ with P.A.=$-24^\circ$, and we
achieved an rms noise of $\sim$0.3 mJy~beam$^{-1}$.  We detected a total
of five sources, whose positions, flux densities and counterparts are
given in Table 1.

In order to obtain higher angular resolution data of the SVS 13 field, we
carried out B configuration observations during 2001 April 20, May 4, and
May 20. The array had then 23 antennas operating at 7 mm. We observed only
the SVS 13 field, using the same phase center and setup as for the D
configuration observations. The absolute calibrator was 1331+305 (adopted
flux density of 1.45 Jy) and the phase calibrator was 0336+323, with a
bootstrapped flux density of 2.8$\pm$0.3 Jy. Because of poor weather
conditions on April 20 and May 20, only the May 4 data were used. Data
were processed using AIPS and cleaned maps were obtained using natural
weighting. The resulting synthesized beam size was
$0\rlap.''18\times0\rlap.''16$ with P.A.=$-65^\circ$, and we achieved an
rms noise of $\sim$0.3 mJy~beam$^{-1}$. Positions, flux densities and
counterparts of the sources detected are given in Table 1.

\section{DISCUSSION}

We detect at 7 mm the sources VLA 2 (MMS3), VLA 7 (IRAS 2A), VLA 10 (IRAS
2B), and VLA 17 (MMS2=SVS 13B) (see Table 1).  These sources were
previously observed at 3.6 and 6 cm by Rodr\'\i guez et al. (1999), where
a discussion on their properties and counterparts can be found. VLA 7, VLA
10, and VLA 2 were further observed at 3.6 cm with higher angular
resolution by Reipurth et al. (2002). Interferometric observations at 3 mm
of VLA 7 and VLA 10 have been recently reported by J{\o}rgensen et al.
(2004).

We also detect SVS 13 at 7 mm, as an unresolved source in the D
configuration, and resolving it in its two components (VLA 4A and VLA 4 B)  
in the B configuration observation (see Table 1). In Figure 1 we compare
the 3.6 cm map observed with the A configuration (Anglada et al. 2000)
with the 7 mm map observed with the B configuration (this paper). Both
maps were obtained with a similar angular resolution of $\sim 0\rlap.''2$.
The positions of the sources in the two maps are in agreement within
$0\rlap.''02$, eliminating, thus, any ambiguity in their identification at
different wavelengths. As can be seen in the figure, at 3.6 cm both
sources present a similar flux density, while at 7 mm VLA 4B is much
stronger than VLA 4A, suggesting that VLA 4B is the dominant source in the
millimeter wavelength range, with a negligible contribution from VLA 4A.

In order to quantitatively confirm this hypothesis it should be verified
that the increase of flux density of VLA 4B over VLA 4A at 7 mm is caused
by dust emission, and not by free-free emission with a steep spectral
index. Usually, the emission at wavelengths longer than a few cm is
dominated by the free-free emission from ionized gas, while the emission
at wavelengths shorter than a few mm is dominated by thermal emission from
dust. Since the wavelength of 7 mm falls in between the centimeter and
millimeter regimes, it should be checked, for each particular object,
which is the nature of the dominant emission at this wavelength. To do
that, we have plotted in Figure 2 the highest angular resolution data
available on SVS 13 in the centimeter and millimeter ranges. The flux
densities of VLA 4B at 3.6 cm and 1.3 cm give a spectral index of
$\alpha=1.36\pm0.26$, which is typical of a thermal ionized jet or a
partially optically thick ionized region. An extrapolation at 7 mm of this
free-free emission yields a flux density of $\sim$1 mJy, much smaller than
the observed value. On the other hand, a fit to the observed data from 3.4
mm to 1.3 mm (that are supposed to trace dust emission) gives a spectral
index $\alpha=2.55\pm0.05$ in the millimeter range, resulting in an
extrapolated flux density at 7 mm in good agreement with the value
observed for VLA 4B (see Table 1 and Fig. 2). Then, we conclude that
free-free emission cannot account for the observed flux density of VLA 4B
at 7 mm, and that a contribution of a different nature, namely, thermal
dust emission, likely from a circumstellar disk surrounding this object,
is required.

In the case of VLA 4A, the overall 3.6 cm, 1.3 cm, and 7 mm data points
can be fitted together with a single spectral index $\alpha=1.25\pm0.20$,
indicating that the observed flux density at 7 mm can be explained as
free-free emission (see Fig. 2). However, we cannot discard a small
contribution from dust emission at 7 mm, since an extrapolation of the 3.6
and 1.3 cm data alone gives a free-free contribution at 7 mm slightly
below the observed flux density of VLA 4A. The difference ($\la$0.8 mJy)
could correspond to a contribution from dust, being it about five times
smaller than in the case of VLA 4B.

In summary, our results indicate that of the two components of the SVS 13
binary system, the source VLA 4B is associated with a much larger amount
of dust than the source VLA 4A. This strongly supports the proposal by
Anglada et al. (2000), who suggested that the millimeter source MMS1 is
the counterpart of the centimeter source VLA 4B. On the basis of the
available optical astrometry, these authors also proposed that VLA 4A is
the counterpart of the visible star SVS 13.

VLA 4B appears as a compact source in our 7 mm map obtained with the B
configuration with an angular resolution of $\sim0\rlap.''2$ (Fig. 1b),
suggesting that the dust emission traced by this source originates in a
compact circumstellar structure, likely a disk, with radius $\la 30$ AU.  
The size of the disk is smaller than that of the typical accretion disks
observed around single T Tauri stars, with radii of 100-150 AU (e.g.,
Dutrey et al.  1996; Wilner et al.  2000), and is more similar to that of
the compact disks observed in the L1551~IRS5 binary system (10 AU;
Rodr\'\i guez et al. 1998), suggesting that they are truncated by the
tidal effects of the companion star. The 7 mm flux density of VLA 4B,
together with the millimeter data shown in Figure 2, can be fitted by a
simple disk model. We assume a geometrically thin, vertically isothermal
disk, characterized by power-law radial dependences of temperature and
surface density, and we adopt the opacity law given by D'Alessio, Calvet,
\& Hartmann (2001).  The data are fitted with a distribution of dust
grains with a maximum size of 1 mm, a temperature distribution
$T(r)=470~(r/\rm AU)^{-0.5}$~K, a surface density $\Sigma(r)=2800~(r/\rm
AU)^{-1}$~g~cm$^{-2}$, implying a disk mass of 0.06 $M_\odot$, for a
radius of 30 AU. Thus, it seems plausible that the emission of VLA 4B is
tracing a protoplanetary disk, since the mass obtained exceeds the minimum
mass required to form a planetary system like the solar one ($\sim
0.01~M_\odot$). Since any dust emission associated with VLA 4A is at least
five times weaker than that of VLA 4B, if a disk was associated with VLA
4A, we expect its mass to be at least five times smaller than that of
the disk associated with VLA 4B. We note that the flux density observed in
the D configuration is larger than the total flux observed in the B
configuration, indicating an additional contribution from an extended
component, likely the infalling envelope.  A fit to the overall spectral
energy distribution, taking into account simultaneously the contributions
of different components (disks+envelope), similarly to what has been done
for L1551 IRS5 (Osorio et al. 2003), would be useful to better constrain
the properties of each component.

Thus, the observational results obtained for SVS 13 imply that it is
feasible that the development of a protoplanetary disk occurs
preferentially in only one of the components of a young close binary
system, with the disk absent or much less significant in the other
component.  In this respect, the case of the SVS 13 binary system appears
to be opposite to the L1551~IRS5 case, where both components of the binary
system are associated with circumstellar disks of dust of comparable
characteristics (Rodr\'\i guez et al. 1998).

These results are in agreement with theoretical simulations (e.g., Bate \&
Bonnell 1997) that show that, depending on the mass ratio of
the components and the specific angular momentum of the system, the
development of circumstellar disks can occur either around a single
component or around both components of the binary. According to Bate \&
Bonnell (1997), a circumstellar disk forms around one of the components of
the binary only if the specific angular momentum of the infalling gas is
greater than the specific orbital angular momentum of that component about
the center of mass of the binary. Thus, if a binary system grows to its
final mass mainly via the accretion of material with low specific angular
momentum, the primary may have a large circumstellar disk, while the
secondary is essentially naked. The systems formed by this method are
expected to be binaries with separations of the order of $\sim 100$ AU,
and they should not be developing a significant circumbinary disk. For
infall with high angular momentum, both components can develop a
circumstellar disk and even a circumbinary disk can be formed.

Under this simplified scheme, the low angular momentum scenario, with a
disk in only one star, appears to correspond to the case of the SVS 13
binary, where the two components are separated by 65 AU in projection, and
where VLA 4A would be the secondary, whereas VLA 4B (the dominant source
at millimeter wavelengths) would be the primary viewed through its
circumstellar disk. On the other hand, L1551~IRS5 apparently corresponds
to a case of accretion of material with higher angular momentum, with both
components associated with a circumstellar disk (Rodr\'\i guez et al.
1998) and probably surrounded by a circumbinary disk (Osorio et al. 2003).

In order to confirm these suggestions, it would be interesting to obtain
the orbital parameters of these systems and to identify which component is
the primary through accurate measurements of absolute proper motions.
Relative proper motions between the two components of the L1551~IRS5
system have been obtained recently (Rodr\'\i guez et al. 2003), and
accurate absolute proper motions have been measured for the T Tau Sa/Sb
system (Loinard, Rodr\'\i guez \& Rodr\'\i guez 2003). At present, no
proper motions are available for the SVS 13 system, but these promising
results obtained for other sources suggest that this goal could be
attainable in a relatively near future, providing a complete test for the
properties of the SVS 13 binary system.

\acknowledgments

We thank P. D'Alessio for providing us the dust opacity code, and L.  
Loinard and an anonymous referee for valuable comments. We thank Z.  
Webster for communicating us her 1.3~mm data before publication,
as well as H. Ungerechts and U.  Lisenfeld for discussion on the IRAM
data. GA, MO and JMT acknowledge support from MCYT grant AYA2002-00376
(including FEDER funds). GA and MO acknowledge support from Junta de
Andaluc\'{\i}a. MO acknowledges support from IAU PGF and AECI. MTB and
RE acknowledge support from MCYT grant AYA2002-00205. RE acknowledges
support from Generalitat de Catalunya grant 2002BEAI400032. LFR
acknowledges support from DGAPA and CONACyT.

\clearpage

\begin{figure}
\plottwo{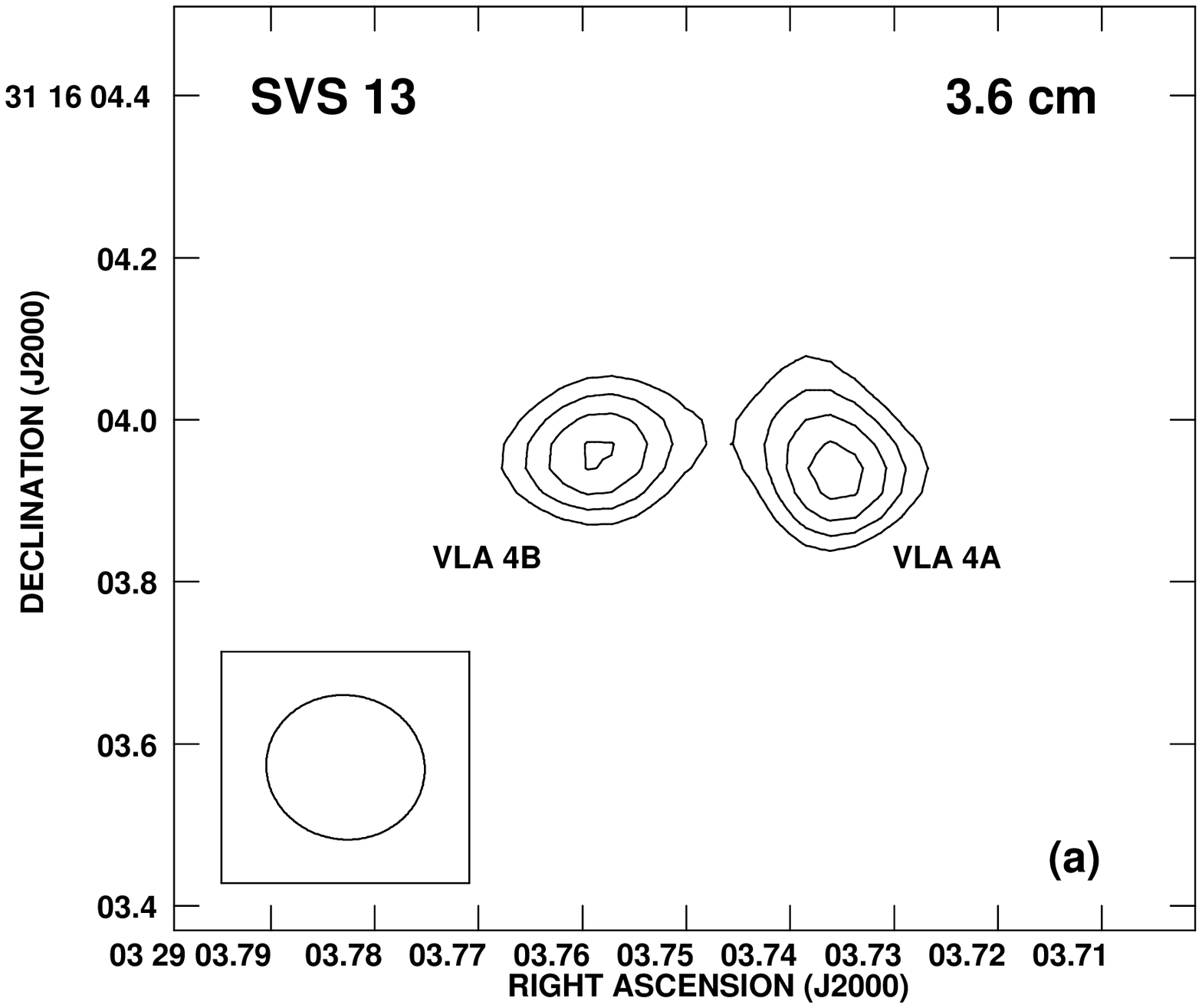}{f1b.eps}
 \caption[fig1.eps]{(a) VLA map (A configuration) at 3.6 cm of SVS~13,
revealing that it is a binary radio source, with component VLA~4A having a
similar flux density as component VLA~4B (Anglada et al. 2000).  
Contour levels are $-3$, 3, 4, 5, and 6 times the rms noise of 14
$\mu$Jy~beam$^{-1}$.  (b) VLA map (B configuration) at 7 mm (this paper).
Contour levels are $-3$, 3, 4, 5, 6, 8 and 10 times the rms noise of 0.3
mJy~beam$^{-1}$. The crosses in this map mark the positions of the 3.6 cm
sources. Note that most of the flux density at this wavelength comes from
the VLA 4B component. The half power contour of the synthesized beam is
shown in the lower left corner of each panel.
 \label{fig1}}
\end{figure}


\begin{figure}
\epsscale{0.85}
\plotone{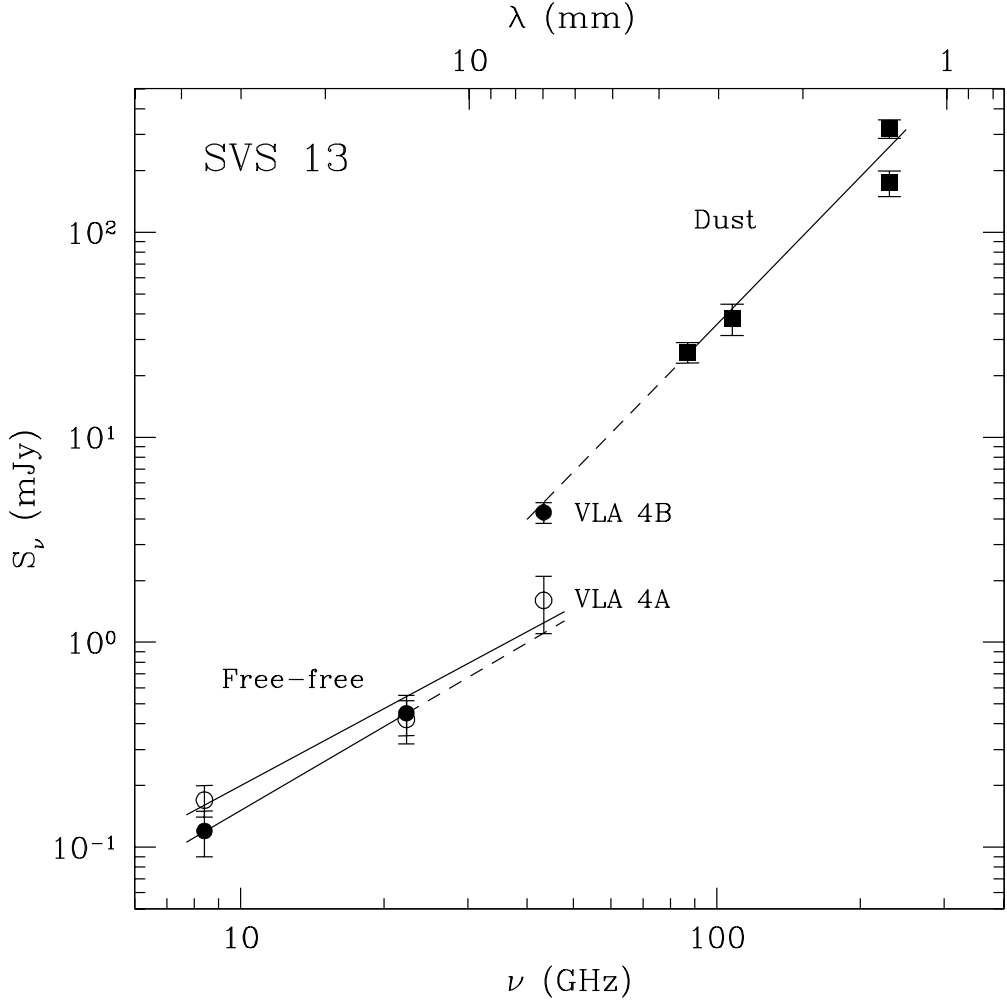}
 \caption[fig2.eps]{Spectrum of SVS~13 in the 3.6 cm to 1.3 mm wavelength
range from data of good angular resolution.  Observed flux densities of
component VLA 4A (empty circles), VLA 4B (filled circles), and of the
overall system (filled squares) are represented.  Solid lines represent
least-square fits to the data points in a given wavelength range, while
dashed lines represent an extrapolation to other wavelengths.  Note that
the observed flux density of VLA 4B at 7 mm is in agreement with the
extrapolation of the fit obtained from the 3.4 to 1.3 mm data (that are
supposed to trace the dust emission), while the flux density of VLA 4A at
7 mm fits the free-free emission. Data points are from VLA observations at
3.6 cm (beam $\simeq 0\rlap.''2$; Anglada et al. 2000), 1.3 cm (beam
$\simeq 0\rlap.''1$; G. Anglada et al., in preparation) and 7 mm (beam
$\simeq 0\rlap.''2$; this paper), Plateau de Bure observations at 3.4 mm
and 1.3 mm (beams $\simeq 3\rlap.''5$ and $1\rlap.''5$;  Bachiller et al.
1998), and BIMA observations at 2.7 mm (beam $\simeq 0\rlap.''6$; Looney
et al. 2000) and 1.3 mm (beam $\simeq 0\rlap.''3$; Z. Webster et al., in
preparation).
 \label{fig2}} 
 \end{figure}


\begin{deluxetable}{lllccl}
\tabletypesize{\small}
\tablewidth{0pt}
\tablecaption{7 mm Continuum Sources \label{tbl1}}
\tablehead{
& \multicolumn{2}{c}{Position\tablenotemark{a}}
& \colhead {$S_\nu$(D config)\tablenotemark{b}}
& \colhead {$S_\nu$(B config)\tablenotemark{c}}
& \colhead {Counterpart\tablenotemark{d}} \\
\cline{2-3}
\colhead{Source}
& \colhead{$\alpha (\rm J2000)$}
&\colhead{$\delta (\rm J2000)$}
& \colhead{(mJy)}
& \colhead {(mJy)}
}
\startdata
VLA 7  &03 28 55.56  &+31 14 37.1  &10.0$\pm0.5$ &\nodata     &IRAS 2A \\
VLA 10 &03 28 57.37  &+31 14 15.9  &5.2$\pm0.3$  &\nodata     &IRAS 2B \\
VLA 2  &03 29 01.962 &+31 15 38.15 &9.0$\pm1.0$  &10.3$\pm1.2$ &MMS3 \\
VLA 17 &03 29 03.072 &+31 15 51.86 &8.5$\pm0.5$  &6.6$\pm0.6$ &MMS2 (SVS 13B) \\
VLA 4A &03 29 03.735 &+31 16 03.95 &10.8$\pm0.6$\tablenotemark{e} 
&1.6$\pm0.5$ &SVS 13 \\
VLA 4B &03 29 03.759 &+31 16 03.94 &10.8$\pm0.6$\tablenotemark{e}  
&4.3$\pm0.5$ &MMS1
\enddata
\tablenotetext{a}{Units of right ascension are hours, minutes, and
seconds, and units of declination are degrees, arcminutes, and
arcseconds. Positions given are from the B configuration data (absolute
positional error is estimated to be $0\farcs05$), except for VLA 7 and
VLA 10, that are from the D configuration data (absolute
positional error is estimated to be $0\farcs2$).}
\tablenotetext{b}{Flux density from the D configuration data.}
\tablenotetext{c}{Flux density from the B configuration data.}
\tablenotetext{d}{See Rodr\'\i guez et al. 1999 and references therein for 
a discussion on the sources VLA 7, VLA 10, VLA 2, and VLA 17 and their 
counterparts. See text for a discussion on VLA 4A and VLA 4B}
\tablenotetext{e}{Total flux density of VLA 4A+VLA 4B. 
The angular resolution in this configuration cannot separate the emission 
of each component, although VLA 4B is likely the dominant one.
}
\end{deluxetable}

\end{document}